# A Novel Light Source Design for Spectral Tuning in Biomedical Imaging

Chandrajit Basu,[(a)] Sebastian Schlangen, Merve Meinhardt-Wollweber, and Bernhard Roth

*Hannover Center for Optical Technologies (HOT), Leibniz University Hannover, Nienburger Strasse 17, Hannover 30167, Germany*

[a)] Email: c.basu@hot.uni-hannover.de

**Abstract:** We propose a novel architecture with a remote phosphor based modular and compact light source in a non-contact dermoscope prototype for skin cancer screening. The spectrum and color temperature of the output light can easily and significantly be changed depending on spectral absorption characteristics of the tissues being imaged. The new system has several advantages compared to state-of-the-art phosphor converted ultra-bright white LEDs, used in a wide range of medical imaging devices, which have a fixed spectrum and color temperature at a given operating point. In particular, the system can more easily be adapted to the requirements originating from different tissues in the human body which have wavelength dependent absorption and reflectivity. This leads to improved contrast for different kinds of imaged tissue components. The concept of such a lighting architecture can be vastly utilized in many other medical imaging devices including endoscopic systems.

## I. INTRODUCTION: Non-contact digital dermoscope

Early detection is a key to successful treatment of melanoma skin cancer. Advanced digital dermoscopic techniques help to detect skin cancer at an early stage and minimize unnecessary excision and biopsy of skin lesions or nevi. The typical commercial contact-type dermoscopes (CTD) used for melanoma screening require the application of index matching gels and human supervision for each and every nevus under test. However, a non-contact digital dermoscope (NCDD) [1] can pave the way for fully automatized and touch-free screening (i.e. without using gels) of melanoma for the whole human body. This can expedite the screening procedure, especially for patients with a large number of nevi. Good image clarity, automated screening with NCDD and software integration of the so called ABCD rule of melanoma detection can improve the speed and quality of melanoma screening in an economically viable manner. The standard ABCD criteria (Asymmetry, Border, Colour, Diameter) are applied to score a suspicious lesion. It has been observed that in case of a non-planar or protruding nevus, the contact plate of the CTD may strongly affect the 3D topology of the nevus and could result in a distorted image of the same. Such distortions can be totally avoided with NCDD and hence more accurate imaging is feasible [1]. Note that pushing the CTD probe on certain skin lesions can be painful for the patient. Naturally, with a remote or non-contact dermoscope, the screening procedure is generally more agreeable for the patient and completely painless.

Contemporary commercial dermoscopes mostly use high power tungsten halogen or multiple white LED based light sources. Our original NCDD prototype incorporates an ultra-bright white LED source, projection optics, imaging optics including crossed polarizers, and a CCD camera [1]. LEDs are efficient, compact and cost effective and hence are preferred over many other light sources these days. LEDs also offer excellent lifetimes in the range of 20,000-50,000 hours which are unmatchable by conventional lamps [2]. However, the spectral limitations of a typical white LED, especially the 'cool white' type, often affect the color rendering and image quality. As image quality is of utmost importance in digital dermoscopy as well as for many other biomedical imaging devices, the design of the corresponding light sources can play a vital role in the overall performance. Also, taking care of the spectra of the light sources can help in the post-processing of the images obtained from typical CCD/CMOS cameras, especially if the post-processing algorithms are based on color based feature extraction. In complex nevi, different kinds of pigments and underlying structures have different optical properties and hence optical probing has an enormous potential for feature based diagnosis. Careful consideration of all these facts led us to the development of an advanced NCDD prototype that utilizes a remote phosphor light source architecture where the output spectra can be varied easily in a very controlled and cost effective manner.

### A. Novel light source application in a non-contact digital dermoscope (NCDD)

The typical phosphor converted (PC) white LEDs are made of a single or multiple blue (~ 460 nm) LED chip(s) with phosphor (e.g. Ce:YAG) coating on top. A part of the blue photons emitted from the chip gets absorbed by the phosphor and down-converted in energy to generate a wide band spectrum ranging from green to red. Hence the residual or unabsorbed blue photons and this down-converted wide spectrum together generate the overall 'white' light output. This is schematically shown in Fig. 1.

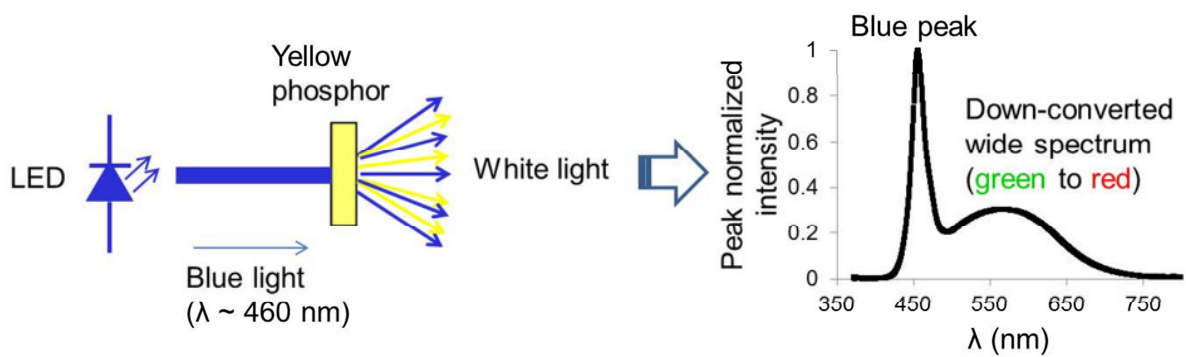

FIG. 1 Schematic of white light generation from a remote phosphor module irradiated by a blue LED.

The same principle of white light generation used in the PC LEDs is utilized in the remote phosphor architecture too, the only difference being a physically separable phosphor module instead of a fixed and 'proximate' phosphor coating. The remote phosphor architecture, a promising approach in the field of general lighting applications, offers many benefits [3]. The physical separation of the blue LED chip and the phosphor module facilitates thermal management. The phosphor module is not exposed to high temperature as on the surface of the blue LED chip and hence it could offer better long term reliability in some cases. Note that remote phosphor modules are reported to have a lifetime in the order of 50,000 hours [4]. On the other hand, remote phosphor design can enhance the output luminous efficacy by up to 60% as compared to standard PC LEDs [3]. This of course involves careful design of the mixing chamber incorporating the blue LED chip and the remote phosphor module. The design of the mixing chamber also plays a significant role in determining the correlated color temperature (CCT) of the output light, for a given spectrum of the blue LED source and the remote phosphor concerned. The nominal CCT values mentioned for commercial remote phosphor modules are valid only for a given blue LED wavelength (e.g. ~ 460 nm) and a reference mixing chamber design. Hence, it should be noted that any variation in the blue source wavelength or mixing chamber design will result in deviations from the nominal conversion efficacy and CCT. However, a detailed treatment on the general technicalities of remote phosphor architecture is beyond the scope of this article.

Discrete RGB LEDs can also be used for good contrast enhancements for skin imaging purposes [5]. However, light field homogenization and full spectral coverage are typically challenging in such cases, unless the skin target is very close to the light sources (e.g. handheld dermoscope) or additional optics for RGB beam homogenization are utilized. Typically each of the RGB LED chips offer a spectral FWHM of about 20-30 nm. Hence they cannot cover the entire region of the white light spectrum obtained from PCLEDs. On the other hand, use of Xenon lamps would consume more space and generate more heat. One has to be careful with possible UV radiation from Xenon lamps as well. Also, the CCT cannot be changed without changing the Xenon lamp itself. In addition, driver requirements are much more complicated for Xenon lamps than in case of LEDs.

Note that the remote phosphor architecture is not limited to the use of blue LEDs as primary sources and blue laser diodes (~ 450 nm) can also be used for high luminance applications [6]. One very interesting application of laser irradiated phosphor source (LIPS) could be in surgical endoscopic systems [7]. With careful design of such a system, CCT of the output white light can be varied in order to see particular tissues with better clarity.

Fig. 2 shows the new NCDD prototype with remote phosphor architecture. Unlike in our previous prototype [1] equipped with an ultra-bright white LED (Luminus CBT-90 White), a blue LED (Luminus CBT-90 Blue) has been used here along with a commercial remote phosphor plate. A mixing chamber is built around the blue LED and the remote phosphor in order to enhance and homogenize the output flux. This new design offers an enormous possibility in terms of spectral tuning by changing the remote phosphor plate itself. One can foresee a situation where a clinician or technician can easily replace one designated phosphor plate with another in order to change the spectra of the light source for different target requirements. Such phosphor plates can be as cheap as < 1$ whereas the ultra-bright white or blue LED can cost around 50-60$ for single pieces from third party vendors/distributors. Hence, for example, replacement of a cool white LED with a warm white one may not be economically viable at all. Commercial remote phosphor plates with a variety of CCT profiles are available for general lighting applications. As a proof of concept, commercial phosphors designed for general lighting applications were sufficient enough for our first tests. However, in future applications with more precise and complex spectral design requirements, phosphor chemists can optimize the composition and tweak the resulting spectra accordingly to a great extent.

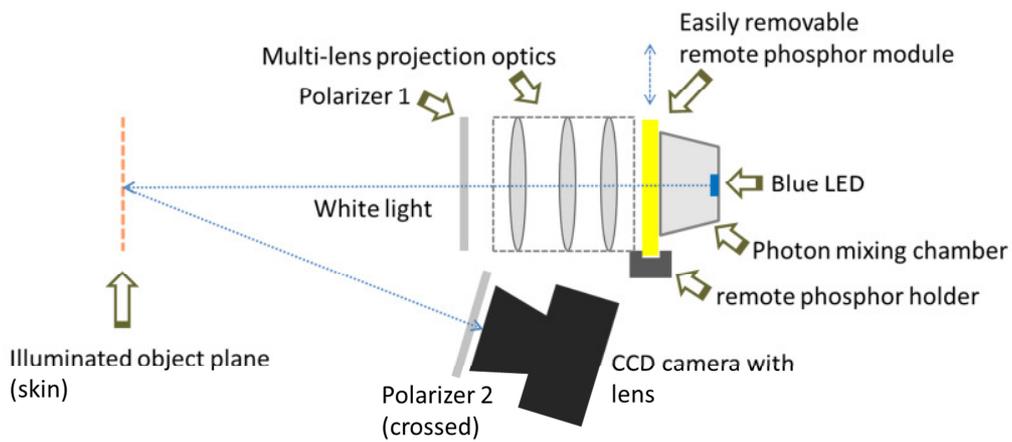

FIG. 2 Schematics of our non-contact digital dermoscope with replaceable remote phosphor based light source

In Fig. 3 the output spectra obtained from two different remote phosphor modules (Ph1 and Ph2) in our prototype setup are compared. The operating current of the blue LED was kept fixed at 5.5 A in both the cases. As obvious, the transmitted spectrum of the blue LED that is used for phosphor excitation in both the cases shows a sharp peak around 460 nm. However, the down-converted output spectra of the phosphors are significantly different. The measured CCT of Ph1 is 6703K whereas a CCT of 3641K is measured with Ph2. In the output white light spectra, normalization to the pump peak shows that the Ph2 output spectrum is twice that of Ph1 at 550 nm and even three times stronger at 625 nm. This can have significant influence on different contrast enhancement mechanisms described in sec. B.

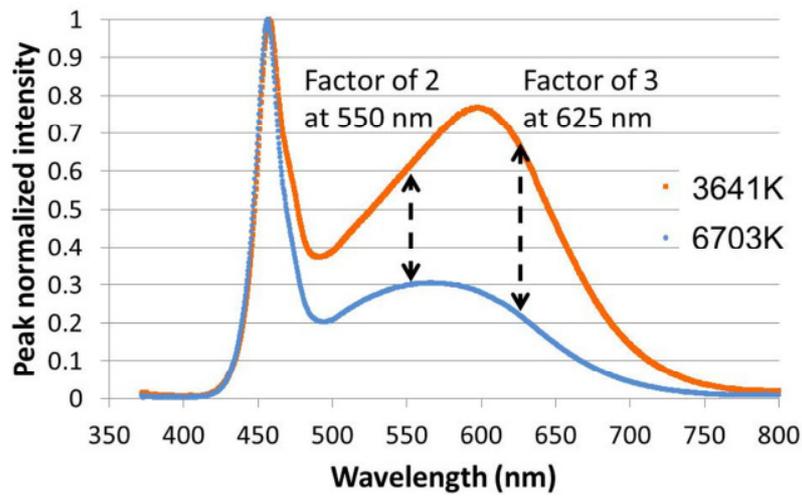

FIG. 3 The spectral comparison of the output white light from two different remote phosphor modules, as measured within an integrating sphere connected to a spectrometer. The CCT values corresponding to the spectra are shown in the legend. The difference in output spectrum for different wavelengths is also indicated.

The CIE 1931 (Commission internationale de l'éclairage) color space chromaticity diagrams in Fig. 4 show the coordinates of the white light output obtained from Ph1 (a) and Ph2 (b). Clearly Ph1 offers a bluish tint (cool white), whereas Ph2 renders a warmer white light, as expected from the spectra in Fig. 3.

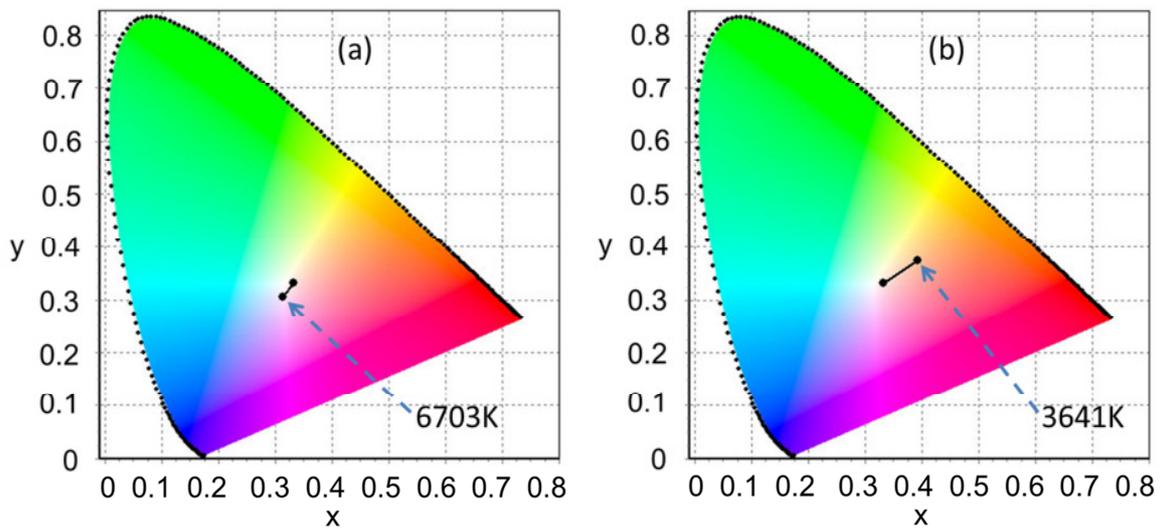

FIG. 4 The CIE color charts showing the color coordinates corresponding to the two different remote phosphors used Ph1 (a) and Ph2 (b). The measured CCT values are pointed by the dashed arrows. The short black line to the measured point is drawn from the CIE 1931 reference white point.

**B. Image contrast enhancement tools and influence of the light source spectra**

Post processing tools can enhance the image contrast of certain components of the nevi images. As far as human skin is concerned, there are several pigments with wavelength dependent absorption characteristics. Hemoglobin and melanin are two of the best-known pigments in human skin. Depending on the nature and degree of a skin disease, the amount and spatial distribution of such pigments can differ significantly from normal skin conditions for the same individual. Hence, the underlying structures of a nevus can be significantly enhanced by using different contrast mechanisms [5] in the post-processing of dermoscopic images. This can uncover features helpful for (early) diagnosis of certain skin diseases or monitor the changes in a nevus over time which is particularly relevant for high risk patients. In all these cases, the doctors would not only see the 'normal' camera images but also can possibly investigate some post-processed enhancements for more structural details which might otherwise go unnoticed. In this article, our discussion will be limited to 'blood contrast enhancement' and 'melanin contrast enhancement' mechanisms only.

i) Blood Contrast Enhancement (BCE)

Blood contrast enhancement [5] is a simple image processing technique used to enhance the visibility of vascular structures (blood vessels) in the papillary dermis of a nevus under investigation. This is favored by the fact that hemoglobin absorbs green light (e.g. 530 nm) much more than red light (e.g. 632 nm). Hence a normalized subtraction of the green channel values from the red channel values can reveal vascularity in great detail as a part of the absorption by melanin in the normal skin background is suppressed in this way. The formula used for BCE, on an image taken by our NCDD prototype with remote phosphor source (Ph2) is schematically shown in Fig. 5. The resulting picture in false colors clearly indicates the enhancement of blood contrast.

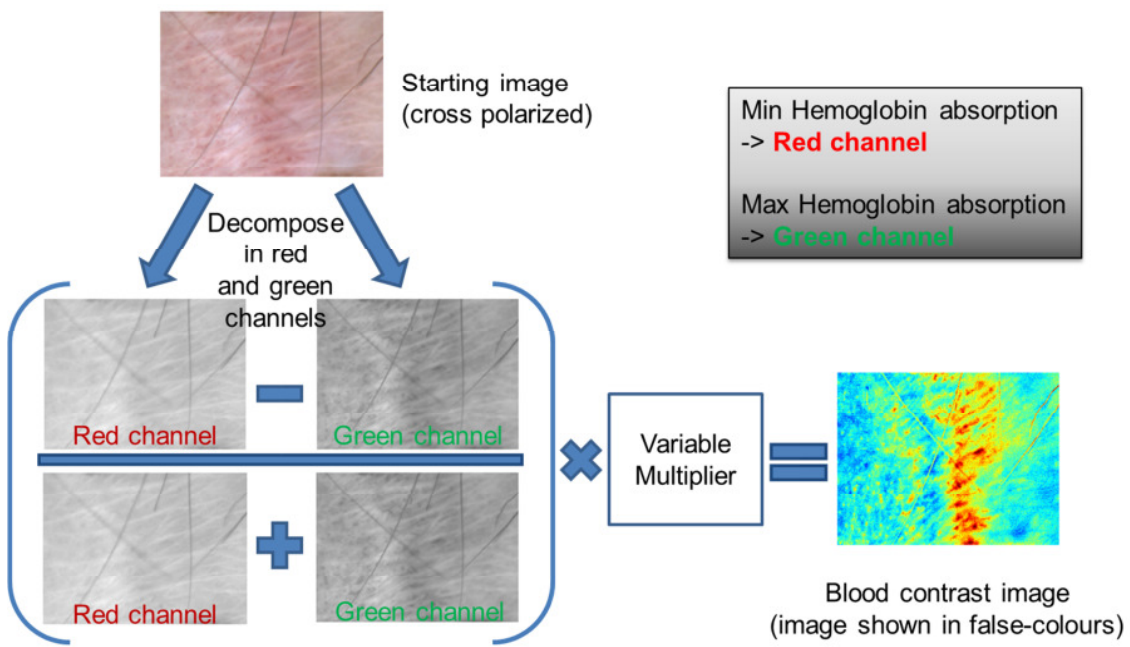

FIG. 5 The BCE mechanism.

The variable multiplier shown in Fig. 5 is basically a gain scaling factor for an 8-bit (0-255) image channel.

ii) Melanin Contrast Enhancement (MCE)

Melanin is the most dominant chromophore in human skin and is normally found in the upper layer of human skin, the epidermis [5, 8]. The pigment is produced by melanocytes, a cell type situated at the junction between epidermis and the underlying dermis and is contained within membranous particles known as melanosomes. In some cases, such as pigmented nevi, melanocytes and melanin are found in the dermis as well. Melanocytic nevi generally have higher melanosome contents than normal skin, which can be used to improve the contrast between the respective skin areas. Besides, data from animal models suggest that benign and malignant melanocytic nevi differ in their depolarization power, which could be due to increased melanin absorption of the tumor [8] or tumor morphology. The red channel has minimum blood contrast and hence is chosen to enhance the melanin contrast by a histogram rescaling mechanism, as shown in Fig. 6. Melanin has an almost exponentially decaying absorption spectrum from UV to NIR [9]. The absorption of both oxygenated hemoglobin (oxyHb) and deoxygenated hemoglobin (deoxyHb), however, is not as even but shows a significant minimum in the range from 600-750 nm [10]. Hence, having a strong intensity level in this range, e.g. at 625 nm, could be helpful while looking for melanin contrast regions in a nevus while minimizing the background absorption from oxyHb and deoxyHb [10]. Note that the choice of a feature based 'diagnostic window' shown in Fig. 6 can be suitably modified depending on the relative contrast between the melanin features and the surroundings (background).

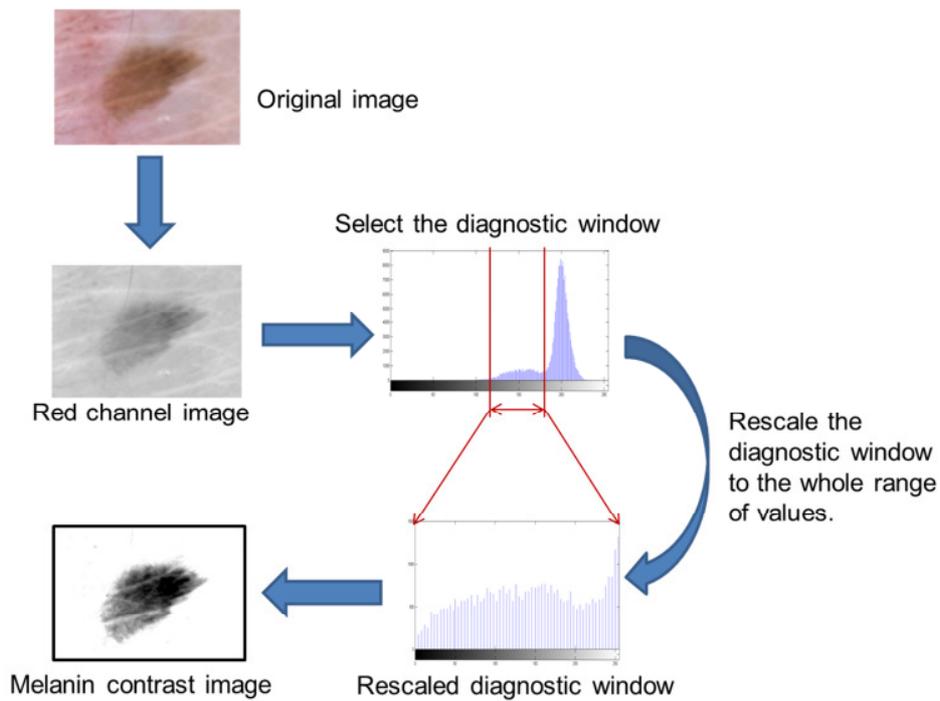

FIG. 6 Algorithm for the MCE mechanism on an image taken by our NCDD system.

Keeping in mind the wavelength specific or say R-G-B channel specific relative absorption levels in the aforesaid contrast mechanisms, as well as the Bayer filters used in typical color CCD/CMOS sensors, it is obvious that by tuning the spectral intensity distribution of the light source one can enhance the contrast and clarity significantly. This could also mean that under certain conditions, the signal to noise ratio (SNR) for particular channels could also be improved from an otherwise 'noisy' output.

**C. Results and Discussion**

The usefulness of our light source architecture is clearly evident in first experiments with our dermoscope prototype, in the context of the aforesaid BCE and MCE mechanisms. Skin patches were illuminated in two different cases using the two remote phosphor modules described before. In Fig. 7, the image processing results for BCE are compared. As argued before, one can clearly see the advantage of having more red and green contents in the light source in order to have greater details on vascularity or blood contrast. Note that the multiplying factor mentioned in Fig. 5 was the same in both cases of the image processing results shown in Fig. 7. The variation of this multiplying factor alone during post processing cannot enhance the contrast beyond a certain limit, unless the red and green contents of the light source are suitably improved too.

The same kind of comparison is shown in Fig. 8 where the MCE mechanism was applied. In this case, a benign nevus with high melanin content was chosen. Again in this case, one can see greater enhancement of the details while using the low CCT phosphor.

The theoretical background and the first results achieved are in good agreement. In co-operation with our collaborating medical partners from the University Medical Center Göttingen (UMG) and the Hannover Medical School (MHH), preclinical trials on more complex nevi samples from various individuals will be pursued in the next steps, using our novel light source based non-contact dermoscope.

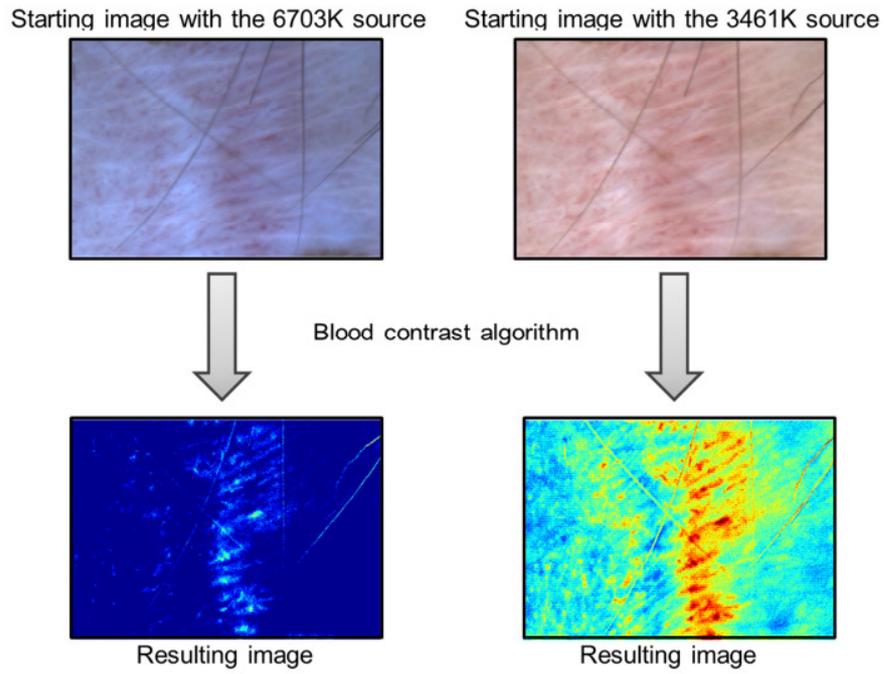

FIG. 7. Comparison of the phosphors for BCE mechanism (left column –Ph1; right column- Ph2).

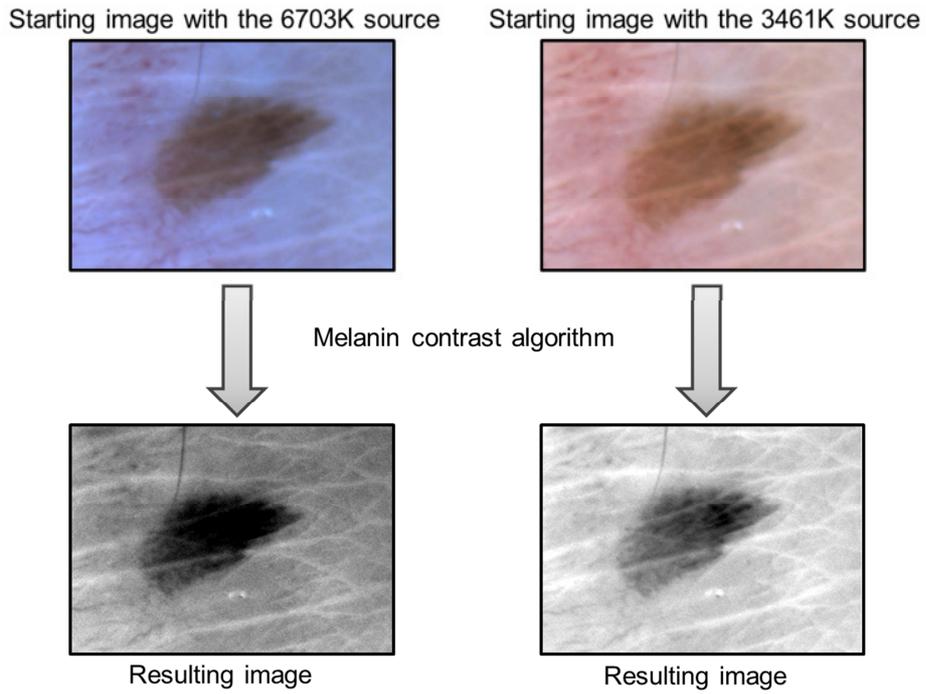

FIG. 8. Comparison of the phosphors for MCE mechanism (left column –Ph1; right column- Ph2).

### D. Conclusion

A novel light source design, with high power blue LED pumped remote phosphor module, for a non-contact digital dermoscope has been developed and experimentally verified. To the best of our knowledge, this is the first time the application of remote phosphor architecture has been demonstrated in a digital dermoscope, achieving easy and cost-effective spectral tuning. This concept can be used in a vast range of other biomedical imaging systems including microscopes. The extraordinary feature of this architecture is that the remote phosphor module is easily replaceable and hence the output spectra can be customized depending on the CCT or spectral requirements, for skin imaging in particular to our case or any other kind of biomedical imaging in general. In applications where variation in the light spectra could be useful, such remote phosphor architecture would possibly be economically more viable than a similar white LED architecture with fixed CCT and spectrum, for changing the white LED is more cost-intensive and often time consuming. Note that although two commercial remote phosphor modules were used in our prototype, in future many other kinds of phosphor fabrication recipes can be utilized to customize the spectra, in accordance with the needs of biomedical imaging. In the context of blood contrast and melanin contrast enhancement mechanisms, the first results of our system are very promising. Further preclinical trials of our non-contact digital dermoscope will follow, in order to study complex nevi (melanoma, psoriasis, acne etc.) with different types of pigmentations and vascularity under customizable spectra from our remote phosphor based light source.